\pdfoutput=1
\documentclass[12pt]{article}
\usepackage{amssymb}
\usepackage{graphicx}

\input{tcilatex}
\begin{document}

\title{The exponentially truncated q-distribution: A generalized
distribution for real complex systems.}
\author{Hari M. Gupta, Jos\'{e} R. Campanha. \\
IGCE - Departamento de F\'{\i}sica, IGCE, UNESP\\
Caixa Postal 178, CEP 13500-970\\
Rio Claro - S\~{a}o Paulo - Brazil}
\maketitle

\begin{abstract}
To know the statistical distribution of a variable is an important problem
in management of resources. Distributions of the power law type are observed
in many real systems. However power law distributions have an infinite
variance and thus can not be used as a standard distribution. Normally
professionals in the area use normal distribution with variable parameters
or some other approximate distribution like Gumbel, Wakeby, or Pareto, which
has limited validity.

Tsallis presented a microscopic theory of power law in the framework of
non-extensive thermodynamics considering long-range interactions or long
memory. In the present work, we consider softing of long-range interactions
or memory and presented a generalized distribution which have finite
variance and can be used as a standard distribution for all real complex
systems with power law behaviour. We applied this distribution for a
financial system, rain precipitation and some geophysical and social
systems. We found a good agreement for entire range in all cases for the
probability density function (pdf) as well as the accumulated probability.
This distribution shows universal\ \ nature of the size limiting in real
systems.
\end{abstract}

\bigskip 

\bigskip 

\textbf{I. Introduction}

\bigskip

\bigskip

To know the statistical distribution of a variable is an important problem
in management. For example, the distribution of the variation of a share
price is important for financial management, while the distribution of the
water flux or water level in a river or rain precipitation is important for
water and flood management.

Recently physicists started to study the natural systems as a whole rather
than in parts [1-6] and are interested in holistic properties of these
systems normally called \textquotedblleft Complex Systems\textquotedblright
. This also include financial, social, biological, economical and
geophysical systems as they have the same characteristics.

Power law scaling [7,8] is observed in many such systems [2,9-29] and it is
now considered an important property of these systems. In general, power law
exists in the central part of the distribution. It deviates from power law
for very small and very large steps.

Long-range interactions and memory effects are present in all real systems
including social and economical sytems [30,31] and are important for a
statistical distribution. Tsallis through non-extensive thermodynamics gives
a microscopic basis for power law [32,33] considering long-range
interactions and long memory effects. The distribution also explain the
initial deviation from power law

Power law have infinite variance which discourage a physical approach and an
unavoidable cut-off is always present. Mantega and Stanley [34] introduced
the truncated L\'{e}vy flight in which the probability of taking a step is
abruptly cut to zero at a certain critical step size. Koponen [35],
gradually truncated the probability distribution from the begining. This
violate power law distribution in central part. Gupta and Campanha [36-38]
proposed the gradually truncated L\'{e}vy flight in which the probability
distribution is cut-off gradually only after a certain critical value. This
distribution have an undesirable discontinuity at the critical length.
Tsallis et. al. [39] consider a cross-over behaviour to explain deviation
from power law for extreme values. This distribution do not explain a sharp
cut-off as observed in many cases [40]. Thus, in absence of a standard
distribution for these systems, in practice, normal distribution with
variable parameters is used. When one is interested in extreme value
distributions, some other distribution like Wakely, Gumbel, exponential,
log-normal etc [41] are used, which are valid only in a particular range and
generally do with provide a physical basis.

A statistical distribution for these systems must have:

(i) finite variance,

(ii) continuous distribution,

(iii) power law in the central part,

(iv) can explain all kind of cut-off from very sharp to very slow for
extreme values,

(v) must have a physical basis for the truncation of the power law.

In the present paper, we propose softing of long-range interactions or long
memory with increase of the variable size under consideration. This avoid
infinite variance. Finally we present a generalised statistical distribution
based on this concept. With the avalaibility of computer programs to
numerically integrate an function, this distribution can be used to
calculate the probability distribution function and accumulated probability
in any range of the distribution with fixed parameters. In Section II, we
present the model and the distribution. In Section III, we present a method
to estimate the parameters of the distribution. In Section IV, we apply this
statistics for many problems in diverse areas and finally in Section V, we
discuss the results.

\bigskip

\bigskip

\bigskip \textbf{II. The model and the statistics}

\bigskip

In 1998, Tsallis [32], presented non-extensive thermodynamics in which he
incorporated long range interactions and long memory effects. He proposed a
generalized definition of entropy $(S_{q})$:

\bigskip 
\begin{equation}
S_{q}=C\frac{1-\sum_{i=1}^{W}p_{i}^{q}}{q-1}
\end{equation}

\[
(\sum_{i=1}^{W}p_{i}=1) 
\]

\bigskip

\noindent where $C$ is a positive constant, and $W$ is the total number of
microscopic possibilities of the system. $q$ is an entropic index, which
plays a central role and is related to long range interactions and long
memory effect in a network. This expression recovers the usual
Boltzmann-Gibbs entropy $(-C\sum_{i=1}^{W}p_{i}\ln p_{i})$ in the limit $%
q\rightarrow 1$, i.e. in short range interactions. In this case, the size
frequency distribution function $N(x)$ is given through

\bigskip

\begin{equation}
\frac{dN(x)}{dx}=-\lambda N(x)
\end{equation}

\bigskip

\noindent where $\lambda $ is a positive constant. $N(x)$ is the frequency
probability of size $x$.

This gives

\bigskip

\begin{equation}
N(x)=N_{0}\exp (-\lambda x)
\end{equation}

\bigskip

In general, the frequency density distribution function N(x) is given
through:

\bigskip

\begin{equation}
\frac{dN(x)}{dx}=-\lambda N^{q}(x)
\end{equation}

\bigskip

\noindent hence

\bigskip

\begin{equation}
N(x)=\frac{N_{0}}{[1+(q-1)\lambda x]^{\frac{1}{q-1}}}
\end{equation}

\bigskip

\noindent where $N_{0}$ is a normalization constant. This expression
recovers the usual Boltzmann distribution in the limit $q\rightarrow 1$ i.e.
in short-range interactions as shown in Euation (3). For $q>1$, this
expression gives power law for relatively large values of the step $x.$

The power law distribution can not continue forever in real systems. It has
to be truncated in some way to avoid infinite variance and have a finite
size.

In order to consider long range departure, Tsallis et. al. [39] assume a
crossover to another type of behavior and modify Equation (4) as

\bigskip

\begin{equation}
\frac{dN(x)}{dx}=-\mu _{r}N^{r}(x)-(\lambda -\mu _{r})N^{q}(x)
\end{equation}

\bigskip

\noindent $\mu _{r}$ is very small compared to $\lambda $. That gives a
crossover between two different power laws (respectively characterized by $q$
and $r$) or from power law to normal distribution within a nonextensive
scenario.

Although cross-over behavior as suggested by Tsallis can avoid an infinite
variance, in the present work, we are looking for another possibility, i.e.,
the truncation of power law due to softing of long-range interaction or
memory which gives finite size in real systems. This is not a cross-over
behavior. We\ consider that entropy factor $q$ decreases with step size $(x)$
due to the softening of long-range interactions or memory effects which
arises because of the physical limitations of the components or the system
itself. Thus q depends on the step size. This is similar as anharmonic terms
are important for calculating potential energy in lattice vibrations.

The size limiting factor is of a very small importance for small steps,
while it is necessary for larger steps. Entropy index $q$ is equal to $1$ in
the absence of long memory or long range interactions. Thus the information
about these interactions are given through $(q-1)$. We consider that this
factor approaches to zero for very large values of $x$. In general, for this:

\bigskip

\begin{equation}
(q(x)-1)=\frac{(q_{0}-1)}{1+\sum\limits_{i}\theta _{i}x^{i}}
\end{equation}

\bigskip

\noindent where $q_{0}$ and $q(x)$ are values of entropy index $q$ for step
size zero and step size $x$ respectively. $\theta _{i}$ and $i$ are
adjustable parameters depending on the softing of long-range interactions or
memory.

To simplify, we propose an exponential decay i.e$.$

\bigskip

\begin{equation}
q(x)-1=(q_{0}-1)\exp (-(\theta x)^{i})
\end{equation}

\bigskip

\noindent where $\theta $ and $i$ show the rate of decrease of the
importance of these interactions with the increase of step size $x.$ The
higher value of $i$ indicates a sharper cut-off. $\theta $ is a scaling
factor for cut-off.

For very large values of $x$, $q(x)$ approaches to $1$ and thus gives normal
distribution as required through central limit theorem. In the present model
the distribution function is given through:

\bigskip

\begin{equation}
N(x)=N_{0}[1+(q_{0}-1)\lambda x\exp (-(\theta x)^{i})]^{-(\exp ((\theta
x)^{i}))/(q_{0}-1)}
\end{equation}

\bigskip

For simplification we replace $(q_{0}-1)\lambda $ by another constant $\beta 
$, and $1/(q_{0}-1)$ through the power exponent $\alpha .$ Finally the
frequency density distribution is given by

\bigskip

\begin{equation}
N(x)=N_{0}[1+\beta x\exp (-(\theta x)^{i})]^{-\alpha \exp ((\theta x)^{i})}
\end{equation}

\bigskip

In Figure 1 we compare $N(x)$ $vs.$ $x$ in present approach through Equation
(10) for very large steps. Under the present model, the gradual truncation
of the power law can be adjusted from very sharp to very slow through the
value of $i$ without interfering in power law behavior in the central part
of the distribution. In cross-over behaviour we can not explain very sharp
cut-off without violating power law in the central part [40].

\bigskip


\includegraphics[height=8.0309cm]{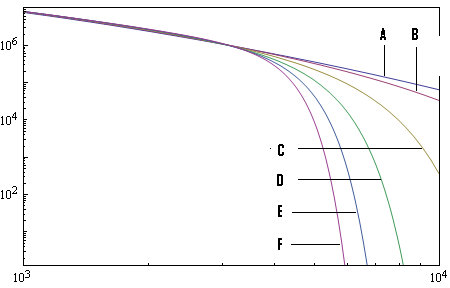}

\textbf{Figure 1 }-- Theoretical distribution of present Model in log-log
scale. We consider: $N_{0}=1.10^{8}$; $\beta =0.0025$; $\alpha =2.0$. Curves
A, B, C, D, E, and F are through considering $i=1/2$ and $\theta =3.10^{-6}$
(Curve A), $i=1$ and $\theta =$ $3.10^{-5}$ (Curve B), $i=2$ and $\theta =$ $%
1.10^{-4}$ (Curve C), $i=3$ and $\theta =$ $1.5.10^{-4}$ (Curve D), $i=4$
and $\theta =$ $1.8.10^{-4}$ (Curve E) and $i=5$ and $\theta =$ $2.10^{-4}$
for Curve F.

\bigskip

In terms of probability density distribution $(p(x))$, the distribution is:

\bigskip

\begin{equation}
p(x)=k[1+\beta x\exp (-(\theta x)^{i})]^{-\alpha \exp ((\theta x)^{i})}
\end{equation}

\bigskip

where k is another normalization constant and equal to $(N_{0}/N_{total})$,
where $N_{total}$ is the total number of observations.

Further

\bigskip

\begin{equation}
\int\limits_{-\infty }^{\infty }p(x)dx=1
\end{equation}

\bigskip

Many times, the maximum frequency is not at $x=0$. In this case we need to
shift the origin to $x_{m}$, where $x_{m}$ is the most frequent value of the
variable $x$. At $x_{m}$ the frequency is maximum, however it is not
necessary a mean value of the variable. In this case the frequency density
distribution is given through:

\bigskip

\begin{equation}
N(x)=N_{0}[1+\beta \left\vert x-x_{m}\right\vert \exp (-(\theta \left\vert
x-x_{m}\right\vert )^{i})]^{-\alpha \exp ((\theta \left\vert
x-x_{m}\right\vert )^{i})}
\end{equation}

\bigskip

The physical mechanisms behind the distribution for $x>x_{m}$ and $x<x_{m}$
may be different and thus the parameters of the distribution may also be
different. Thus the two cases must be treated separately.

For $\theta =0$, the present distribution turn out to be Tsallis
distribution [33,39] as is expected.

The accumulated probability in between $a$ and $b$ is given through

\bigskip

\begin{equation}
P(a\leq x\leq b)=\int\limits_{a}^{b}p(x)dx
\end{equation}

\bigskip

The value of $p(x)$ at $x=0$ must be carefully studied, particularly when $%
x_{m}=0$. In many cases it may be a discrete number and include many cases
apart from the mechanism under discussion. For example, in financial market,
the variation zero in the price of a particular share also include cases in
which it is not at all traded along with cases in which it is traded but
with zero variation. Thus in these cases $p(0)$ should be separately
estimated through

\bigskip

\begin{equation}
p(0)=\frac{Events\text{ \ }x=0}{Total\text{ \ }events}
\end{equation}

\bigskip

and thus

\bigskip

\begin{equation}
\int\limits_{-\infty }^{\infty }p(x)dx=p(0)+\int\limits_{-\infty
}^{0^{-}}p(x)dx+\int\limits_{0^{+}}^{\infty }p(x)dx=1
\end{equation}

\bigskip

where $0^{-}$ and $0^{+}$ are respectively the values of $x$ below and above
zero. We define $N(x)$ as events in the range $((x-\frac{\Delta x}{2})<x<(x+%
\frac{\Delta x}{2}))$ divided by $\Delta x$ and thus eliminate $p(0)$ when $%
x=\frac{\Delta x}{2}$.

There is very little mass for extreme values of x and thus it is difficult
to compare a theoretical distribution with available data. The technique
known as a Zipf plot [38] is very important in this case. Suppose we ordered
our observations from largest to smallest so that the index $i$ is the rank
of $x_{i}$. Then

\bigskip

\begin{equation}
i=N\int\limits_{x_{i}}^{\infty }p(x)dx
\end{equation}

\bigskip

Thus the rank is simply a transformation of the accumulative distribution
function. The empirical accumulated probability above $x_{i}$ is its rank $i$
divided by total number of observations. The accumulated probability
accentuates the upper tail of the distribution and therefore makes it easier
to detect the deviations in the extreme tails from the theoretical
predictions of a particular distribution.

\bigskip 

\bigskip \textbf{III. Estimation of parameters.}

\bigskip

We use the following steps to estimate the parameters of the distribution.
Let $(x_{1},x_{2},...x_{N})$ be the set of $N$ observations of a random
variable $x$ for which the probability density function is $p(x)$. We select
a proper bin size $\Delta x$ and make a frequency table $f(x)$ $vs$ $x$. $%
f(x)$ gives the number of observations in between $(x-\frac{\Delta x}{2})$
and $(x+\frac{\Delta x}{2})$. From this table we observe the value of $x_{m}$
i.e. the value of $x$, for which we have the maximum value of the frequency $%
(N_{0})$. We separate the observations in two groups, one for $x>x_{m}$ and
the other for $x<x_{m}$. Each group may have different values of parameters
because of different mechanisms in two cases.

The frequency density function $N(x)$ is given by:

\bigskip

\begin{equation}
N(x)=\left\vert \frac{f(x)}{\Delta x}\right\vert _{x}
\end{equation}

\bigskip

In the case of extreme values of x, we have very little mass i.e. very few
observations and thus, we have many zeros in $f(x)\_vs\_x$ table for the
extreme values because of the limited and random nature of the observations.
Thus to make a physical significance of observations, we increase the value
of the interval $\Delta x$ for the extreme values to avoid zero values of $%
N(x)$.

For small steps, cut-off parameters are of negligible importance. We
therefore put $\theta =0$ and $i=0$ for inicially $50\%$ steps and estimate $%
\alpha $ and $\beta $. $N_{0}$ and $x_{m}$ are estimated through frequency
tables. Knowing these parameters, we estimate $\theta $ and $i$ for the best
fit for the entire curve. Some time it may be necessary to re-estimate $%
\alpha $ in this stage.

\bigskip

\bigskip

\bigskip \textbf{IV. Applications.}

\bigskip

Now we apply this model to describe the distribution of a parameterin some
geophysical complex systems of interest

\bigskip

(A) Water level of a river:

\bigskip

For water, flood and agriculture management, it is extremely important to
know the the distribution of water level in a river of interest. It is
therefore regularly registered by water management department. In the
present case we took the water level in Paran\'{a} River, one of the
important river in Brazil, at the S\~{a}o Paulo-Paran\'{a} border. The water
level is measured daily by the Agencia Nacional de Agua and can be obtained
at the site www.ana.gov.br. We analize water level in the period of $1^{st}$
of January of 1964 to $30^{th}$ of June of 2005, in total having 15,067
observations. This river receives water from many sources and the water
level depends on rainfall at different places at different times and thus
present a complex system with long term memory and long-range interactions.

Through frequency distribution of the the empirical data, we observe $\ $%
maximum frequency density $N_{0}=10,032$ days/m at the heigth of 2.87 m $%
(x_{m})$. For $x\geq x_{m}$, we found $\alpha =1.66,$ $\beta =1.25$, $\theta
=0.185$ and $i=2.85$ for best fit. For $x\leq x_{m}$, we found $\alpha =3.2,$
$\beta =0.54$, $\theta =0.5$ and $i=9.0$

In Figure 2a we compare $log(frequency)$ $vs$ $water$ $level(\mathbf{a})$
distribution. Plotting log(frequency) we can compare cases of even very
small frequency at extremely higher water levels. In Figure 2b, we compare
accumulated probability distribution density above water level \textbf{a} $%
P(x>\mathbf{a})$ $vs$ $\mathbf{a}$. The agreement is good for the entire
curve including extreme cases up to four orders of magnitude in accumulated
probability density as well as in frequency density distribution.

\bigskip


\includegraphics[height=8.0309cm]{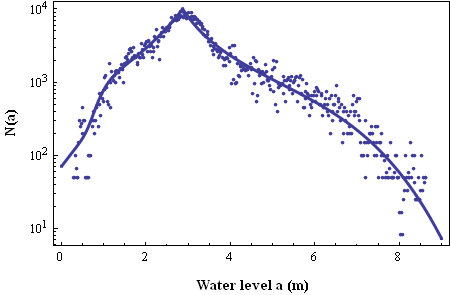}

\begin{center}
\textbf{Figure 2a - }Frequency density vs water level (m) in semi log scale.
The continuous line is through present model. The dotted points are empirical

\bigskip
\end{center}


\includegraphics[height=8.0309cm]{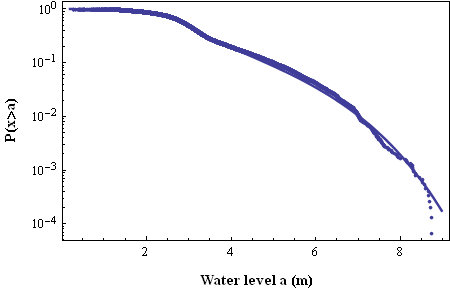}

\begin{center}
\textbf{Figure 2b} - Accumulated probability density above water level $\ 
\mathbf{a}$ $(P(x>\mathbf{a}))$ $vs$ $water$ $level(\mathbf{a)}$ in semi log
scale. The dotted points are empirical, while the continuous line is through
present model

\bigskip
\end{center}

\bigskip (B) Water flux in a river:

\bigskip

Another problem of interest in water management is the distribution of water
flux in a river at a particular point of interest. We have 5,428
observations in the period of $23^{th}$ of October of 1969 to $31^{th}$ of
August of 1984 in Paran\'{a} river at S\~{a}o Paulo-Paran\'{a} border.
Through frequency distribution of the observations data, we find $N_{0}=24.4$
days.seg./m$^{3}$ at $x_{m}=155$ $m^{3}/s$. For $x\geq x_{m}$, we found $%
\alpha =0.84,$ $\beta =0.013$, $\theta =0.00177$ and $i=3.0$. For $x\leq
x_{m}$, we found $\alpha =0.8,$ $\beta =0.045$. Due to the small number of
observations on this side, we did not observe any truncation of q-values and
thus consider $\theta =0$ and $i=0$. In Figure 3a we cmpare $log(frequency)$ 
$vs$ $water\_flux$ while in Figure 3b we compare accumulated probability
distribution density $P(x>\mathbf{a})$ above $water$ $flux(\mathbf{a})$ $vs$ 
$water$ $flux(\mathbf{a})$. Again the agreement is good throughout the curve
including extreme values

\bigskip


\includegraphics[height=8.0309cm]{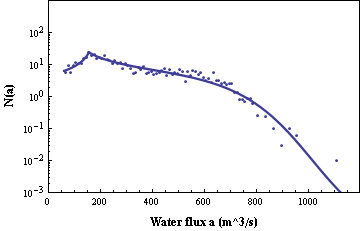}

\bigskip 

\begin{center}
\textbf{Figure 3a} - Frequency density vs water flux $(m^{3}/s)$ in semi log
scale

\bigskip
\end{center}


\includegraphics[height=8.0309cm]{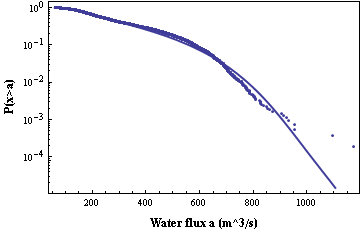}

\bigskip 

\begin{center}
\textbf{Figure 3b} - Accumulated probability density for water flux above 
\textbf{a} $(P(x>\mathbf{a}))$ vs water flux \textbf{a} in semi log scale.
The dotted points are empirical, while the continuous line is through
present model
\end{center}

\bigskip

\bigskip (C) Rain precipitation

\bigskip

To compare the distribution of rain precipitation, an importante problem for
agriculture and water control, we took a time series of daily rain
precipitation at Campinas city in S\~{a}o Paulo State, Brazil, in the period
of 1950 to 1980 at station prefix D4-044. These data were obtained from the
Departamento de Aguas e Energia Eletrica of S\~{a}o Paulo State and are
avalaible at http://www.sigrh.sp.gov.br. In total we have 21,549
observations.

In case of rain precipitation, probability of precipitation zero is a
singular point as it also includes days when there is no rain at all. We are
only considering days when there is rain precipitation more than zero. We
have $N_{0}=598$ days/mm for $x_{m}=1.0$ $mm$. The values of the parameters
are $\alpha =1.7,$ $\beta =0.074$, $\theta =0.08$ and $i=0.12$. The number
of days when there is no rain precipitation is 15.558 days, thereby giving $%
p(0)=0.722$.

In Figure 4a, we plotted $log(frequency)$ $vs$ $rain$ $precipitation$. In
Figure 4b we plotted accumulated probability density distribution $P(x>%
\mathbf{a})$ for rain precipitation above \textbf{a} versus rain
precipitation.\textbf{a} Again the agreement is good.

\bigskip


\includegraphics[height=8.0309cm]{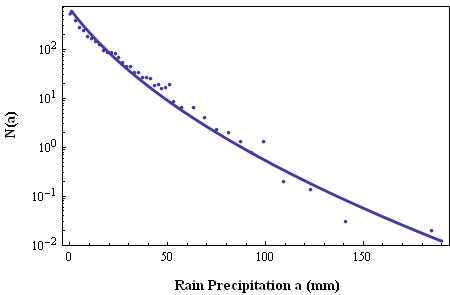}

\begin{center}
\textbf{Figure 4a }- Frequency density vs rain precipitation in mm.
\end{center}

\bigskip


\includegraphics[height=8.0309cm]{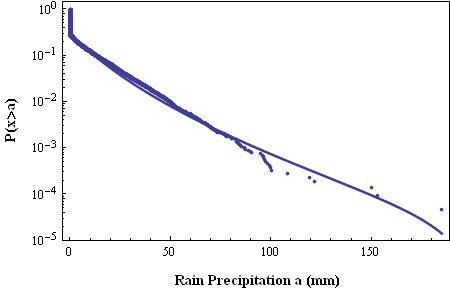}

\begin{center}
\textbf{Figure 4b} - Accumulated probability density for rain precipitation
above a $(P(x>a))$ vs rain precipitation a in semi log scale. The dotted
points are empirical, while a continuous line is through present model.
\end{center}

\bigskip

\bigskip \bigskip (D) Financial Systems:\bigskip Variation of an economical
index.

\bigskip

It has been shown recently that the variation of a share price in high
frequency limit i.e. variation per minute is given through power law [36].
However for extreme values the predicted variation by power law is much more
than what is observed and it must be truncated in some way or other. In the
present case we took variation of the price of the share of Banco do Brasil,
the biggest semi-government bank of Brazil. We consider variation per
minute, i.e. in high frequency range. The period is from $1^{st}$ of July of
2004 to $30^{th}$ of June of 2007 in total of 329,489 observations and
furnished by IBOVESPA - S\~{a}o Paulo. In this case the probability of the
variation zero is a singular point as it also includes all those minutes,
when the share is not at all traded along with minutes when the share is
traded but with zero variation.

For positive variation we have frequency density $N_{0}=29,640$ at $%
x_{m}=0.5.10^{-3}$ $\%$. The values of the parameters\ for this side is $%
\alpha =5.5,$ $\beta =0.197.$ We did not observe the effect of gradual
truncation in this period and so we put $\theta =0$ and $i=0$. For negative
side we found $N_{0}=29,640$ at $x_{m}=-0.5\ast 10^{-3}$ $\%$. The values of
the parameters\ for this side are $\alpha =4.0,$ $\beta =0.3325$, $\theta
=1.10^{-7}$ and $i=0.12$. This means that for this side, the effect of the
truncation although small, still is necessary. In Figure 5a, we shown
log(frequency) $vs$ percentage variation in unit of $10^{-3}$ The agreement
is good. In Figure 5b we only compare the accumulated probability for the
10,000 highest variations in frequency density as they are most important.
Again we found a good agreement.

\bigskip


\includegraphics[height=8.0309cm]{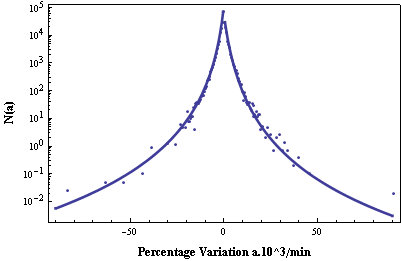}

\begin{center}
\textbf{Figure 5a - }Frequency density vs. percentage variation $(.10^{3})$%
\textbf{\ }in share price/min.

\bigskip
\end{center}


\includegraphics[height=8.0309cm]{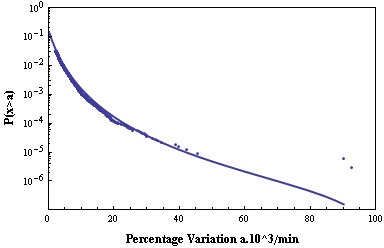}

\begin{center}
\textbf{Figure 5b }- Accumulated probability density for share price
variation above \textbf{a} $(P(x>\mathbf{a}))$ vs. share price variation for
extreme values in semi log scale. The dotted points are empirical, while a
continuous line is through present model..

\bigskip
\end{center}

\bigskip \bigskip (E) Distribution of the sun spots.

\bigskip

The number of sun spots per month is a very old index and is avalaible from
1611. This index measures the magnetic activity in the sun. Sun spot number
data can be obtained from National Geophysical Data Center in Boulder,
Colorado and is avalaible at the site http://www.ngdc.noaa.gov.

In the Figure 6a we show the distribution of monthly sun spots from 1749 to
2007 and compare with present model with parameters $N_{0}=58.4$ month$^{2}/$%
sun spot for $x_{m}=2.5,$ $\alpha =0.8,$ $\beta =0.04$, $\theta =0.0075$ and 
$i=1.9$. In Figure 6b we compare the accumulated probability density for $%
x\geq a$ versus $a$. The agreement is good in both cases.

\begin{center}
\bigskip \bigskip 

\includegraphics[height=8.0309cm]{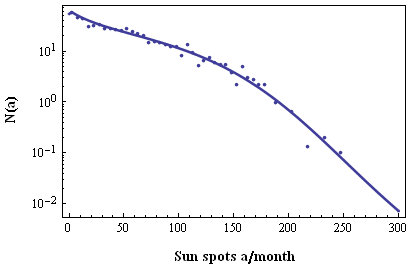}

\textbf{Figure 6a }- Frequency density vs. sun spots/month

\bigskip
\end{center}


\includegraphics[height=8.0309cm]{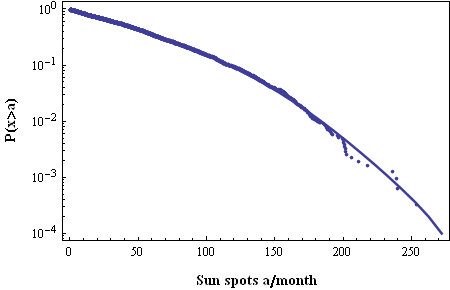}

\begin{center}
\textbf{Figure 6b }- Accumulated probability density for sun spots/month
more than \textbf{a} $(P(x>\mathbf{a}))$ vs. sun spots/month (\textbf{a}) in
semi log scale. The dotted points are empirical, while a continuous line is
through present model..
\end{center}

\bigskip

\textbf{V. Discussion}

\bigskip

In the present work, we presented a statistical distribution considering
that the entropic index ($q-1$), which gives information about long range
interactions and/or memory effects, decreases exponentially with size of the
variable. This distribution automatically gives a power law in the central
part and deviates for very small and very large values of the variable as
normally is observed in most of the complex systems. It gives finite
variance as required through central limit theorem. In the present work we
applied this model for various geophysical and financial systems, and found
a good agreemnt in all cases. We tried this model in some other cases like
citation index of scientists and marks distribution in an entrance
examination [40], citation index of scientific publications [42] where also
we obtain a good agreement for eight order of magnitude. In certain cases,
due to limited observations, we could not estimate the values of gradually
truncation parameters and thus consider them equal to zero.

This distribution present an statistics for complex systems which is valid
for entire range and can be used by geophysical and financial professionals.
Thus it elliminate the necessity to use distribution with variable
parameters or an approximate distribution which is valid only in a limited
range. It has a strong physical basis and we have shown a good agreement up
to four orders of magnitude or more. Thus it provide a confiable standard
distribution for these systems. This model also present an universal nature
of the truncation process in the distribution of a parameter of a complex
system obeying power law.

\medskip

.

\end{document}